\documentclass[preprint, 12pt]{aastex}

\def\citett#1{\citeauthor{#1} \citeyear{#1}}

\begin{document}
\title{Shot cutoff timescales in different spectral states of Cygnus X-1}
\author{Hua Feng\altaffilmark{1}, Shuang-Nan Zhang\altaffilmark{2,3,4,5},
and Ti-Pei Li\altaffilmark{2,4}}

\altaffiltext{1}{Department of Engineering Physics and Center for
Astrophysics, Tsinghua University, Beijing 100084, China;
fenghua01@mails.tsinghua.edu.cn}

\altaffiltext{2}{Physics Department and Center for Astrophysics,
Tsinghua University, Beijing 100084, China}

\altaffiltext{3}{Physics Department, University of Alabama in
Huntsville, Huntsville, AL 35899, USA}

\altaffiltext{4}{Laboratory for Particle Astrophysics, Institute
of High Energy Physics, Chinese Academy of Sciences, Beijing
100039, China}

\altaffiltext{5}{Space Science Laboratory, NASA Marshall Space
Flight Center, SD50, Huntsville, AL 35812, USA}

\shortauthors{Feng, Zhang \& Li}

\shorttitle{Cyg X-1 Shot Timescales and State Transitions}

\begin{abstract}
We investigate the shot cutoff timescale evolutions during
different spectral states in the black hole binary Cygnus X-1 with
the recently proposed $w$ spectral analysis technique. For low
energy shots, their cutoff timescale decreases from the low state
to intermediate state, and to high state monotonically. However in
the high energy range, where shots are believed to be reprocessed
in the hot corona, timescales are almost the same for different
spectral states. A linear correlation is found between the energy
spectrum photon index and the shot cutoff timescale at low
energies. Both narrow and broad iron K$\alpha$ lines are
distinguished from timing analysis, providing dynamical evidence
that broad and narrow iron lines originate respectively at small
and large radii of the accretion disk. Possible mechanisms for
state transitions and accretion flow geometry in accreting stellar
mass black holes are discussed.
\end{abstract}

\keywords{black hole physics --- accretion, accretion disks ---
X-rays: binaries --- stars: individual (Cygnus X-1)}

\section{Introduction}\label{sec:intro}

The Galactic X-ray source Cyg X-1 is one of the brightest stellar
mass black holes and is the first object classified as a black
hole X-ray binary (BHXB) \citep[for reviews of BHXBs see
e.g.][]{tan95,mcc04}. BHXBs are known to exhibit five different
X-ray spectral states, i.e., the quiescent, low/hard,
intermediate, high/soft and very high state. In the canonical
black hole system Cyg X-1, only three of them have been seen,
i.e., the common low/hard state with a power-law spectrum of
photon index $\Gamma \sim$ 1.4--1.9 with a characteristic cutoff
around 100 keV \citep{don92,dov98,pot03a}, and the high/soft state
with a power-law index $\Gamma$ $\sim$ 2.2--2.7 which often
extends up to 500 keV without a cutoff
\citep{zha97b,cui97,cui02,pot03b}, and the intermediate state with
properties between the low and high states \citep{bel96}.

The understanding of state transitions is usually associated with
different properties of the accretion flow. Most models relate
states with different accretion rate $\dot m$, e.g., the
advection-dominated accretion flow (ADAF) model
\citep{esi97,esi98}. In the ADAF models of \citet{esi98},
different states are described by different $\dot m$ and
transition radii, i.e., $r_{tr}\gtrsim 30 R_s$ (Schwarzschild
Radius) for the low state and $r_{tr}\approx3 R_s$ for the high
state. However, recently \citet{lu04} have shown that
$r_{tr}\approx6 R_s$ for a smoothly matched Shakura-Sunyaev disk
and ADAF. Alternatively, \citet{zha97a} proposed that the high and
low states might correspond to either a prograde or a retrograde
disk around a Kerr black hole; a temporary disk reversal leads to
transitions in a wind accretion system such as Cyg X-1. In this
model, the ratio between the inner accretion disk radii in low and
high states is 2--3. \citet{you01} suggested that a surface
ionization change would cause the state transition; the ratio of
the accretion power dissipated in the corona to that dissipated in
the disk determines the spectral state; in this model the inner
accretion disk radius remains unchanged. \citet{hom01} argued that
the other parameter must exist to trigger state changes apart from
$\dot m$ and might be a relative size of a Comptonizing region.
Therefore, the state transitions are not yet understood
completely.

The iron K$\alpha$ emission has been used as an effective probe to
study relativistic effects around black hole systems, because it
is one of the most prominent spectral features generated very
close to the innermost stable circular orbit (ISCO) of the black
hole \citep{rey97,wan00,wil01,fab02,rey03}. Among all Galactic
BHXBs, Cyg X-1 produces one of the brightest iron K$\alpha$ line
emissions, with both narrow and broad lines discovered by a
variety of instruments (for \textit{ASCA} see \citett{ebi96}; for
\textit{BeppoSAX} see \citett{fro01}; for \textit{Chandra} see
\citett{mil02} and \citett{sch02}). Current understanding of Fe
lines is provided mainly from energy spectral analysis; limited
dynamical evidence has been available.

Recently \citet[hereafter Paper I]{fen04} proposed a new
statistics $w$ to detect shot timescales very sensitively, which
is defined as the differential coefficient of the mean absolute
difference of the time series $x\{{i,\tau}\}$, with $i$ being the
index and $\tau$ being the time resolution:
\begin{displaymath}
W(\tau) = \frac{\overline{|x(i+1;\tau)-x(i;\tau)|}}{\tau},
\;\;\;w(\tau) = -\frac{dW(\tau)}{d\tau},
\end{displaymath}
where the averaging process is on the index $i$. By selecting
different time resolution $\tau$ through binning on the original
time series, a spectrum of $w$ is obtained. In practice it is
necessary to eliminate the digital phase effect of binning and
subtract the background component; please refer to Paper I for
details. From simulations we know the $w$ spectrum is very
sensitive to shot width, but insensitive to shot profile,
amplitude and periodicity; the timescale in $w$ spectrum refers to
the width of a pulse signal rather than the 1/frequency in the
Fourier concept.

With this technique black hole binaries and neutron star binaries
are well distinguished depending upon if there is a cutoff at
small timescales in the $w$ spectrum. For a black hole binary such
as Cyg X-1, its $w$ spectrum presents a sharp cutoff at the small
timescale, which can be measured accurately and is thus defined as
the ``shot cutoff timescale''. A brief introduction of this
technique for energy dependent analysis of shot cutoff timescale
is presented in \S\ \ref{sec:diss}. In this paper, we make further
analysis on Cyg X-1 data with the $w$ spectrum, by presenting its
shot timescales at different energies in its three spectral
states, as well as its shot timescale evolution within each state
when its flux and energy spectral index change. We also report the
discovery of both broad and narrow Fe line features from Cyg X-1's
shot timescale analysis, providing dynamical evidence for the
origin of the Fe lines.

\section{Observations and data analysis}

We use observations from the Proportional Counter Array (PCA)
onboard the \textit{Rossi X-ray Timing Explorer} (\textit{RXTE}).
Data sets used in this paper include observations of two low
states, two intermediate states, and two high states (for details
see Table \ref{data}). The data were screened with pointing offset
$<$0.02\arcdeg\ and elevation angle $>$10\arcdeg. The light curves
were recorded with a time resolution of 1 ms for 1998-Dec-25 and 4
ms for others, divided into 8 energy channels in binned data mode
at low energies and event encoded mode at high energies. Energy
spectra are extracted from Standard-2 files including only top
layer events, and fitted in the \textit{XSPEC} 11.3 \citep{arn01}.
When calculating the $w$ spectra, every segment contains 50,000
bins.

\begin{figure}
 \centering
 \epsscale{0.8}
\plotone{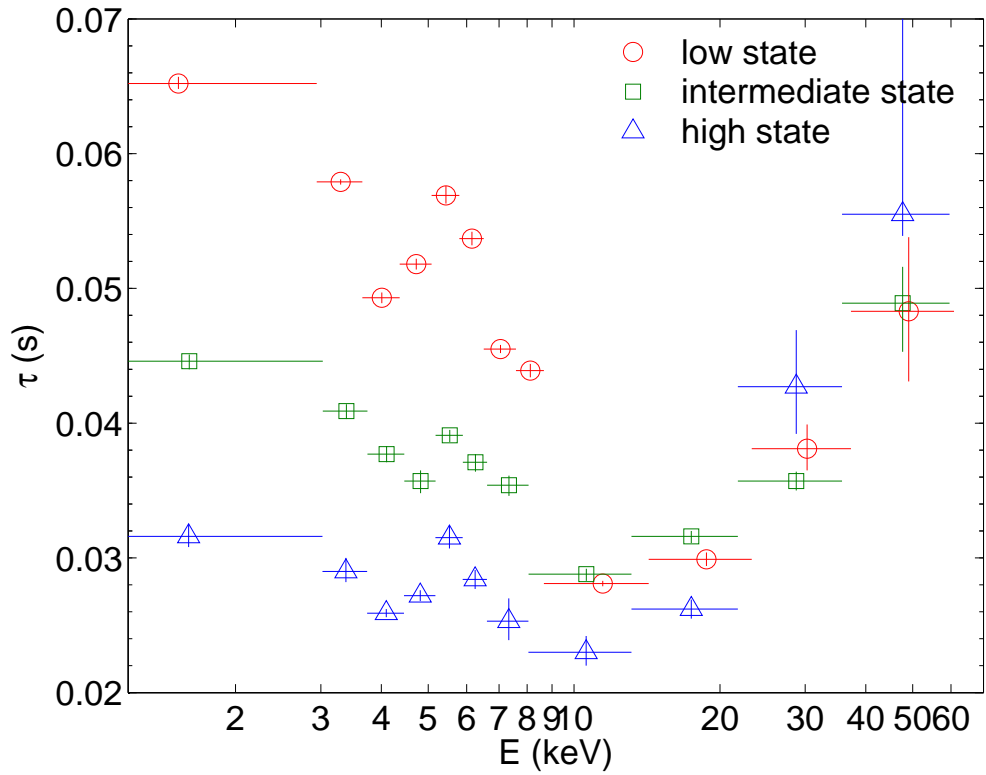}
 \caption{Shot cutoff timescales versus energy in three states of Cyg
X-1. At energies below the ``turning point'' of around 10 keV,
shot timescale increases from high state to intermediate state and
then to low state, respectively. However above the tuning energy,
shot timescales are almost the same for the three states.}
 \label{ec3}
\end{figure}

In Figure \ref{ec3}, the energy dependent shot cutoff timescales
$\tau$ in the low, intermediate and high states of Cyg X-1 are
presented respectively. The profile contains two parts, a
descending part at low energies and an ascending part at high
energies, separated by a ``turning point'' located around 10 keV
(for details see Paper I). It is shown clearly that below the
``turning point'' (the descending part), cutoff timescales
decrease when the spectral state becomes high/soft. However above
the ``turning point'', timescales in three states are almost the
same.

\begin{figure}
 \centering
 \epsscale{0.8}
\plotone{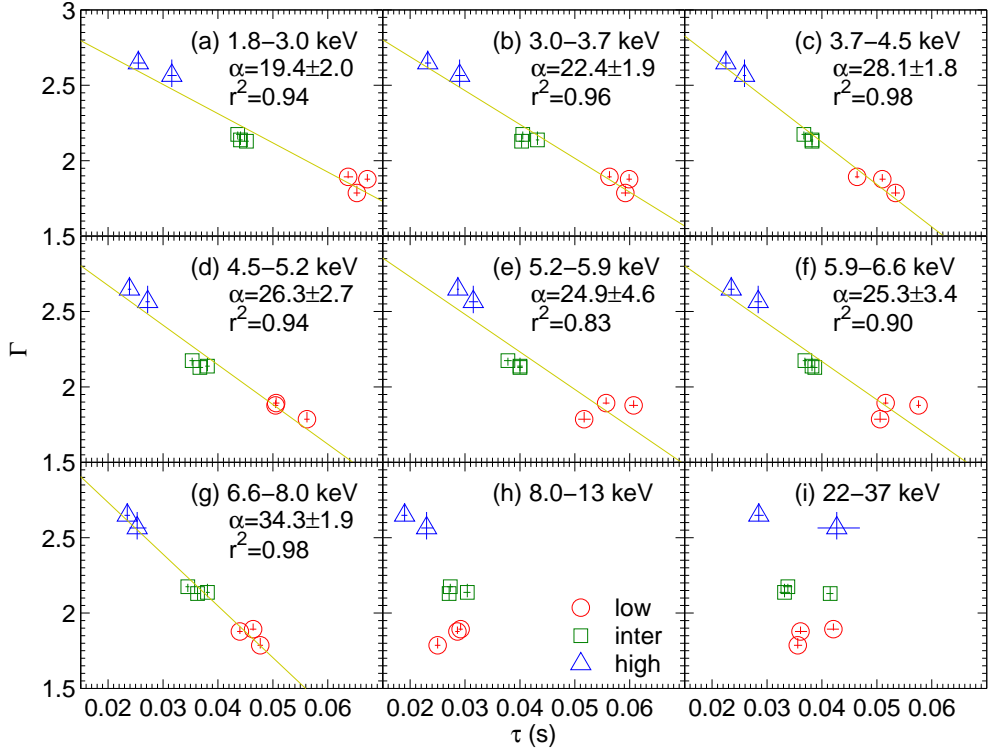}
 \caption{Photon index $\Gamma$ versus shot cutoff timescale $\tau$
in different spectral states of Cyg X-1 at different energy bands.
An inversely proportional relationship is found between $\Gamma$
and $\tau$ in low energy bands from (a) to (g). However in higher
energy bands (h) and (i), no correlation can be found. In the Fe
K$\alpha$ bands (e) and (f), the correlation at low states is
weak. The solid line is the best-fitted line with the slope
$\alpha$, standard error and correlation coefficient labelled. It
seems in low energy bands $\alpha$ increases with increasing
energy except in the Fe line region.}
 \label{gamma_tau}
\end{figure}

To study the inner-state shot timescale evolution as well as the
state transitions, power-law spectrum photon indices $\Gamma$ and
shot cutoff timescales $\tau$ at different states and different
energy bands are shown in Figure \ref{gamma_tau}. Every energy
spectrum is fitted by a multi-color disk model \citep{mit84,mak86}
plus a power-law model with a reflection component \citep{mag95},
i.e., diskbb$+$pexrav, in the energy range from 3 to 30 keV. It is
obviously shown that at low energy bands from panels (a) to (g),
$\Gamma$ is inversely proportional to $\tau$, with a solid line
indicating the best-fitted line with slope, standard error and
correlation coefficient labelled. While at energies larger than
the ``turning point'' in panels (h) and (i), there is no
correlation between $\tau$ and $\Gamma$. Also in the Fe K$\alpha$
band in panels (e) and (f), the correlation at low state is
somewhat weak. The best-fitted line slope $\alpha$ at the low
energy bands seems to increase with energy increasing except in
the Fe line band.

From a large number of shot cutoff timescales versus energy, two
typical instances, one in the intermediate state (due to its
intermediate counting rates and photon index) and the other in a
low state, are selected and shown in Figure \ref{fe}. The one in
the intermediate state has a broad Fe peak and the other in the
low state contains both broad and narrow Fe lines.

\section{Discussion}\label{sec:diss}
With the help of the $w$ spectrum, shot cutoff timescales can be
sensitively detected. An energy dependent analysis may be used to
constrain the different spatial scales of the accreting system
(see Paper I). In this work, we study the evolutions of the shot
timescales at different energies: a long-term evolution
corresponding to state transitions and a short-term evolution
corresponding to the inner-state variations.

In the scope of our discussion, we only consider the shot model as
\citet{pou99}, in which the shot timescale is related to the
Keplerian timescale, and therefore corresponds to the spatial
scale of the accretion flow .

It is useful to explain clearly the meaning of the ``shot cutoff
timescale''. For a monochromatic shot, the $w$ spectrum may be
used to detect its width distribution. A complete study of shot
behaviors should be carried out in a three dimensional space
including timescale distributions at different energies. The
cutoff timescale is thus a kind of characteristic timescale in
this distribution, corresponding to roughly the shortest shot
width of all shots at the given energy. For instance, in our
consideration of shots modeled by \citet{pou99}, the shot cutoff
timescale at a certain energy means the smallest radius of the
disk or the corona where the energy shot is generated. Please
refer to Paper I for further explanations and discussions on the
meaning and significance of the ``shot cutoff timescale''.

The shot cutoff timescale versus energy for an accreting black
hole system contains three components, the descending and
ascending components separated by a ``turning point'', and a broad
peak component about 6 keV. The descending part is possibly
originated from emissions in the disk, with higher energy shot
located at inner radius causing a decreasing of timescale with
increasing energy. The ascending part is related to emissions
(properly speaking, reprocessed emissions) from the corona; shots
are Compton up-scattered to increase energy as well as timescale.
In this scenario, the ``turning point'' is thought to correspond
to the innermost region of the accretion disk, where radiations
are most variable. The broad peak around 6 keV is caused by the Fe
K$\alpha$ fluorescence deduced from the relevant energy spectral
analysis.

We also want to make it explicit that the measured timescale at
the Fe emission line energy only indicates the behavior of the
line shot rather than the continuum shot. The time lag of a shot
in Cyg X-1 is much smaller than the shot timescale \citep{neg01}.
Therefore, the measured light curve profiles are overlapped for
line shot and continuum shot, with the overlapped width
approximately equal to the line shot width. Since the $w$ spectrum
is only sensitive to shot width rather than shot profile, the
detected timescale at Fe line energy only reflects the width of
the Fe line shot.

Based on these understandings, we interpret the results in the
following parts separately.

\subsection{Disk truncation radius in different states}
From the descending part in Figure \ref{ec3}, the difference
between timescales in the low state $\tau_l(E)$ with that in the
high state $\tau_h(E)$ decreases with the energy $E$ until the
``turning point'', at which, $\tau_l(E_t)=0.028$ s and
$\tau_h(E_t)=0.023$ s for these two observations respectively.
Emissions from the ``turning point'' are thought being generated
from the innermost region of the disk. If the shot timescale is
related to the Keplerian timescale \citep[e.g., in the magnetic
flare model by][]{pou99}, i.e., $r\propto\tau^{2/3}$, the
innermost radius ratio for low and high states is
$r_l/r_h=(\tau_l/\tau_h)^{2/3}=1.14$. However this result is
significantly different from the ADAF estimate of \citet{esi98},
in which the ratio between the transition radii for the low and
high states is larger than 10. This ratio is marginally consistent
with the ADAF model of \citet{lu04}, or the reversal disk model of
\citet{zha97a}. The ionization model proposed by \citet{you01}
seems to explain the result more naturally, in which the disk will
always extend to the ISCO; different states correspond to
different ionization parameters and surface densities determined
by the ratio of energy dissipated in the corona to the disk.

\subsection{Corona}
It is reasonable to assume that the ascending part in Figure
\ref{ec3} is dominated by corona (reprocessed) emissions due to
the high photon energies, and partially affected by the disk
reflection component. The Comptonization in the corona will
naturally show a positive correlation between energy and
timescale, with multiple scattering expanding shot width and
increasing shot energy simultaneously
\citep[e.g.,][]{sun80,tit94}. The extended corona model
\citep{kaz97} with a $r^{-1}$ radial density distribution predicts
shot width grows with energy. However the auto-correlation
function (ACF) analysis gives an opposite result, with ACF width
decreasing with energy from 2 to 40 keV \citep{mac00}. Our $w$
spectrum is sensitive to real shot pulse rather than the
variability of the thermal emissions (see detailed comparisons in
Paper I), and is thus more suitable for shot analysis. In fact,
the predicted width-energy correlation from the extended corona
model is verified in the shot cutoff timescale versus energy at
energies above $\sim$10 keV for Cyg X-1, as shown in Figure 1.

We make Monte-Carlo simulations of the extended corona model to
compare with our results. In this model, the corona consists of an
uniform-density inner core and a $r^{-1}$-density outer shell,
with the same density at the boundary between the inner core and
outer shell. The electron temperature $kT$ and optical depth
$\tau$ are fixed in simulation from energy spectral fitting, as
$kT$ of 46, 21 and 26 keV, $\tau$ of 2.87, 1.53 and 0.61 for the
low, intermediate and high state, respectively (the low state
values are fitted from the HEXTE data; the intermediate and high
state values are taken from \citett{mac02}). Therefore, the
best-fitted inner core radius $R_c$ is 11, 7 and 30 $R_s$, and the
inner core density $\rho_c$ is 2.0, 1.9 and 0.3 $10^{16}$cm$^{-3}$
for the low, intermediate and high state, respectively; we adopt
the mass as 10$M_\sun$ \citep{her95,kub98} and the inclination
angle as 35\arcdeg\ \citep{gie86} for Cyg X-1.

Our results indicate that in the low state, the corona is smaller
and denser than the high state; the corona configuration in the
low and intermediate state is similar. This conclusion contradicts
the unified accretion model of \citet{esi97}. We would like to
mention that the uniform-density corona model
\citep[e.g.][]{pay80,tit94} provides a much worse fit to our
results compared to the extended corona model.

\subsection{Evolutions of the accretion disk within and between states}
From Figure \ref{gamma_tau}, a tight linear correlation is found
between photon index $\Gamma$ and shot cutoff timescale $\tau$ at
low energies, in particular at energy bands without Fe K$\alpha$
contamination. This correlation exists between the three states as
well as within every individual state. Because the tightest
correlations appear at low energy bands outside of the Fe line
region, of which emissions are thought to originate from the disk
without reflection, this might suggest that the accretion disk
evolves in the same way from the low/hard state to the high/soft
state: for shots with the same energy, a steeper power-law
corresponds linearly to a shorter cutoff timescale, and thus a
smaller accretion disk radius at which the shots are generated.
Therefore this linear correlation may place tight constraints to
any theoretical model for state transitions in black hole
binaries.

\subsection{Fe K$\alpha$ behavior}

\begin{figure}
 \centering
 \epsscale{0.8}
\plotone{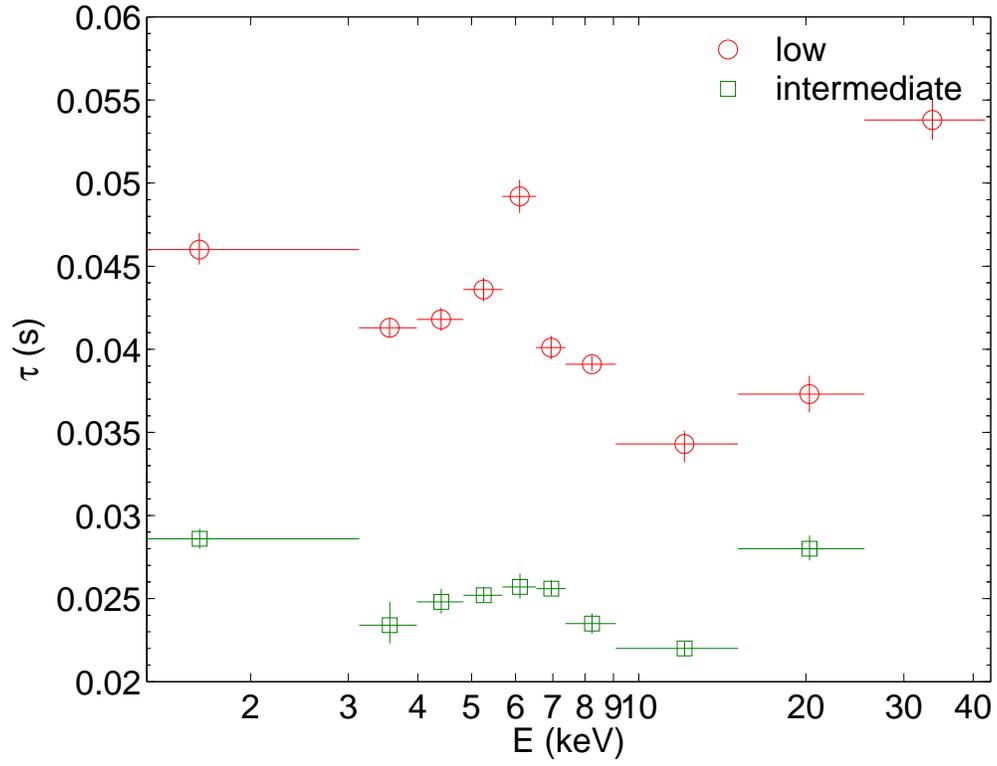}
 \caption{Shot cutoff timescales versus energy in two observations
of Cyg X-1. The intermediate state contains a broad Fe K$\alpha$
line. The low state contains a narrow line above a weak broad
line.}
 \label{fe}
\end{figure}

Timescale-broadening of both broad line (Case I) and narrow line
(Case II) is revealed in our results (Figure \ref{fe}), with Case
I in an intermediate state, and both Case I and Case II in a low
state. Case I is possibly caused by light bending
\citep{fab03,min04}. For Case II, the increased timescale
$\Delta\tau$ possibly due to light travelling time can be inferred
from the irradiating region scale $r$ and the disk inclination $i$
as $\Delta\tau=r(1+\sin i)/c$, denoting the crossing time between
the center and the far side of the irradiated disk by a point
source located in the center. From Figure \ref{fe}, due to the
minimum timescale from corona is the ``turning point'',
$\Delta\tau\lesssim 0.015$ s, thus corresponding to the irradiated
region $r\lesssim100R_s$. If considering the corona with a scale
$R_c$ covering the disk, the irradiated region will be larger, as
evidence that the narrow Fe line is produced at large radii from
the central black hole.

\acknowledgments We thank Dr. J. M. Wang and Dr. Y. X. Feng for
helpful discussions, and Dr. X. M. Hua for the Comptonization
codes. The anonymous referee is appreciated for insightful
comments and suggestions. This work was supported in part by the
Special Funds for Major State Basic Science Research Projects of
the Ministry of Science and Technology, by the Directional
Research Project on High Energy Astrophysics of the Chinese
Academy of Sciences, and by the National Natural Science
Foundation of China through grant 10233030. S. N. Z. also
acknowledges support by NASA's Marshall Space Flight Center and
Long Term Space Astrophysics Program.

\clearpage

\begin{deluxetable}{cccl}
 \tablecaption{Involved Cyg X-1 data sets$^\dag$ form \textit{RXTE} PCA \label{data}}
 \tablehead{\colhead{Obs ID} & \colhead{Obs date} &
 \colhead{State} & \colhead{Shown in}}
 \startdata
10412-01-01-00 & 1996-05-23 & intermediate & Fig. \ref{ec3} \& \ref{gamma_tau} \\
10512-01-08-00 & 1996-06-17 & high & Fig. \ref{ec3} \& \ref{gamma_tau} \\
10512-01-09-01 & 1996-06-18 & high & Fig. \ref{gamma_tau}\\
40100-01-01-00 & 1998-12-25 & low & Fig. \ref{ec3} \& \ref{gamma_tau} \\
60089-02-01-01 & 2001-09-23 & intermediate & Fig. \ref{fe} \\
60089-02-02-04 & 2001-10-28 & low & Fig. \ref{fe} \\
 \enddata
\tablenotetext{\dag}{Some of these data sets have been analyzed
previously \citep[e.g.,][]{che00a,che00b,liu04,fen04}. In this
paper the light curves and energy spectra were re-extracted
separately with the latest version of HEASOFT and calibration
database.}
\end{deluxetable}

\clearpage

\end{document}